# Verifying Policy Enforcers*

Oliviero Riganelli[1], Daniela Micucci[1], Leonardo Mariani[1], and Yliès Falcone[2]

[1] University of Milano Bicocca Viale Sarca 336, IT-20126 Milan, Italy
email: {riganelli,micucci,mariani}@disco.unimib.it
[2] Univ. Grenoble Alpes, CNRS, Inria, Grenoble INP, LIG, 38000 Grenoble, France
email: ylies.falcone@univ-grenoble-alpes.fr

**Abstract.** Policy enforcers are sophisticated runtime components that can prevent failures by enforcing the correct behavior of the software. While a single enforcer can be easily designed focusing only on the behavior of the application that must be monitored, the effect of multiple enforcers that enforce different policies might be hard to predict. So far, mechanisms to resolve interferences between enforcers have been based on priority mechanisms and heuristics. Although these methods provide a mechanism to take decisions when multiple enforcers try to affect the execution at a same time, they do not guarantee the lack of interference on the global behavior of the system.

In this paper we present a verification strategy that can be exploited to discover interferences between sets of enforcers and thus safely identify a-priori the enforcers that can co-exist at run-time. In our evaluation, we experimented our verification method with several policy enforcers for Android and discovered some incompatibilities.

## 1 Introduction

Software ecosystems provide new challenges to verification and validation techniques. An ecosystem is typically composed of a marketplace, where software applications are published and made available to the public, application developers, who implement and share their applications through the marketplace, and customers, who search, download, and use the applications in the marketplace [22]. Notable examples of marketplaces are Android's Google Play, and Apple's App Store.

Marketplaces represent useful channels that enable direct communication between developers and customers. However, marketplaces also expose customers to several threats. In fact, it is extremely hard to control the quality of every application published on a marketplace, and thus marketplaces end up containing a number of unreliable, unsafe, and unstable applications [5, 7, 36, 34, 37].

In addition to enriching marketplaces with advanced mechanisms to check the quality of the published applications [25], customers can exploit richer execution environments to protect themselves from the execution of untrusted software applications. In this context, the *policy enforcement* technology provides mechanisms to automatically detect the violations of correctness policies and enforce the correct behavior at runtime. These solutions have been already experienced in several contexts, including the Android environment [8, 16, 32, 24, 12].

---

* The final publication will be available at Springer's website

While activating a single enforcer that can guarantee that a given policy is satisfied is not problematic, there might be issues when multiple policies should be guaranteed simultaneously. In fact, the policy enforcers might *interfere* one with the other, introducing unexpected behaviors whose effect might be even worse than the result produced by the monitored application without the enforcers.

So far, the problem of interfering enforcers has been addressed using *priority mechanisms* that can disambiguate at runtime which enforcer to execute when multiple enforcers need to react to a same event in different ways [9]. While these mechanisms can be effective in some cases, they suffer from three limitations:

- *Direct Interference*: Priority mechanisms are not adequate when enforcers have to modify a same execution at multiple points to guarantee a sound behavior of the application. For instance, an enforcer may automatically release the microphone acquired by an app when the app is paused, but may also need to acquire and assign the microphone back to the same app when the execution of the app is resumed. If the conflict resolution policy let the enforcer modify the execution only once, for instance because a second higher priority (interfering) enforcer prevents the acquisition of the microphone by the first enforcer, the resulting execution may produce highly undesirable results. For instance, the app may fail once the execution is resumed because the formerly acquired microphone is not available anymore.
- *Indirect Interference*: Enforcers might interfere even if not impacting on exactly the same events. For instance, two enforcers may independently act on two dependent resources (e.g., the Android media recorder and the microphone) producing an interference. In fact, releasing the Android media recorder also releases the microphone which cannot be used anymore unless it is acquired again. Thus an enforcer monitoring the usage of the media recorder while enforcing certain policies may interfere with an enforcer doing the same for the microphone.
- *Late Detection of Interferences*: Even when interferences are on the same events, priority mechanisms are heuristic solutions that operate at runtime to guarantee that at least one policy is correctly enforced. On the contrary, an a-priori analysis of possible interferences allows to detect the interferences in advance. This information is useful both to users, who might be prohibited to activate incompatible sets of enforcers, and to developers, who could redesign the enforcers in such a way that all the policies are correctly enforced.

In this paper, we present an interference detection strategy that overcomes the aforementioned problems for enforcers defined as *edit automata*, which is the most used formalism to define the behavior of enforcers that can manipulate executions [27]. The analysis is designed for enforcers that prevent applications from misusing the resources available in their execution environment. In order to apply the analysis, the *lifecyle* of the applications, which is the same for every application, and the *usage protocol* of the resources, which does not depend on the specific app that uses the resources, must be known.

Note that if a set of enforcers that monitor interactions with some resources does not interfere according to our analysis, they can be activated together regardless of the specific applications that are executed, because the application lifecycle and the usage policies are always the same.

We applied the analysis to 25 enforcers designed to guarantee that Android apps use resources appropriately, and we have been able to verify the compatibility of the enforcers, also discovering some interferences.

The paper is organized as follows. Section 2 provides background definitions. Section 3 introduces a motivating case that is used to illustrate the analysis. Section 4 presents the analysis for interference detection. Section 5 presents our experience with Android. Finally, Sections 6 and 7 discuss related work and provide final remarks, respectively.

## 2   Background

This section defines three concepts that are exploited in the paper to define the interference analysis: policies, edit automata, and I/O automata.

### 2.1   Policy

Let $\Sigma$ denote a finite set of observable program actions $a$. An *execution* $\sigma$ is a finite or infinite non-empty sequence of actions $a_1; a_2; \ldots; a_n$. The notation $\sigma[i]$ denotes the $i$-th action in the sequence. The notation $\sigma[\ldots i]$ represents the prefix of $\sigma$ until the $i$-th actions, and $|\sigma|$ represents the length of the sequence. The symbol $\epsilon$ denotes the empty sequence, that is, an execution with no actions.

$\Sigma^*$ is the set of all finite sequences, while $\Sigma^\omega$ is the of infinite sequences. Finally $\Sigma^\infty$ is the set of all the sequences (both finite and infinite).

Given a set of executions $\chi \subseteq \Sigma^\infty$, a *policy* is a predicate $P$ on $\chi$. A policy $P$ is satisfied by a set of executions $\chi$ if and only if $P(\chi)$ evaluates to *true*.

*Policy Example.* The Android framework includes the `MediaPlayer` API for the playback of audio/video files and streams. To use the media player in their applications, developers must obtain an instance of a `MediaPlayer` by invoking the class method `create()`. The acquired media player instance can be released by invoking the instance method `release()`.

According to the Android documentation, to make the media player available to other applications and to avoid resource leaks, the usage of the `MediaPlayer` should be governed by the following policy:

> Policy 1: "*if you are using a `MediaPlayer` and your activity receives a call to `onStop`, you must release the `MediaPlayer`.*" [2]

For the purpose of the analysis presented in this paper, we represent policies as CTL formulas [6]. We use CTL because it is the language supported by UP-AAL [10], which is the verification tool that we used in the empirical evaluation. However, our properties express linear-time behaviour, thus they could also be expressed with LTL. For example, *Policy 1* can be defined as

$$AG(\texttt{MediaPlayer.create} \Rightarrow$$
$$AXA[\neg\texttt{Activity.onStop}\ W(\texttt{MediaPlayer.release})])$$

which states that "once the `MediaPlayer` is created, the `Activity` can be stopped only after the `MediaPlayer`' has been released".

## 2.2   Edit Automata

Edit automata can be used to describe how policies can be enforced at runtime [26]. An edit automaton is an abstract machine that specifies how an execution is transformed by inserting and suppressing actions. More formally, an edit automaton $A_E$ is a tuple $< \Sigma, Q, q_0, \delta >$ where:

- $\Sigma$ is a finite or countably infinite set of *actions*;
- $Q$ is a finite or countably infinite set of *states*;
- $q_0$ is the *initial state*;
- $\delta : Q \times \Sigma \to Q \times \Sigma^*$ is the *transition function* that maps a state and an action to the new state reached by the automaton and the finite or empty sequence of actions emitted by the automaton that is indicated by the second component in the returned pair. When the emitted action is the same as the accepted action, the automaton does not affect the execution. In the other cases, the actions that are actually executed are influenced by the edit automaton. Action suppression is represented with the empty sequence.

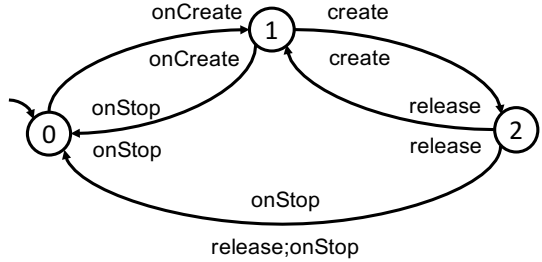

**Fig. 1.** Edit Automaton $EA_{p1}$ enforcing *Policy 1*

*Example of edit automata.* Figure 1 shows the $EA_{p1}$ edit automaton, which can enforce policy *Policy 1* at runtime. The symbol above a transition indicates the input symbol accepted by the automaton, while the sequence below a transition indicates the output sequence emitted by the automaton when the input sequence is accepted.

In the initial state (state 0), $EA_{p1}$ accepts a call to the `onCreate` callback method, which represents the creation of an activity[3]. The creation of an activity causes a transition from state 0 to state 1 in the model. When the activity is destroyed, the `onStop` callback is emitted and the model moves back to state 0. In these cases, the execution is never modified, that is, the transition always emits the accepted action.

State 1 also accepts a call to the `create` method, which returns an instance of the `MediaPlayer`. This case corresponds to the app starting to use the `MediaPlayer`. It causes a transition to state 2 in the model, while the execution is left unaltered by the edit automaton. State 2 is the state that can detect the violation of the resource usage policy, if any. In fact, if the `onStop` callback method is detected, the application is paused without releasing the `MediaPlayer`. The automaton fixes the execution by intercepting the call to `onStop` and emitting the sequence `release;onStop` (transition from state 2 to state 0), which forces the release of the `MediaPlayer`. On the contrary, if `release` is emitted, *Policy 1* is satisfied and the model does not change the execution.

### 2.3   I/O Automata

An input/output automaton is a labeled state machine typically used for modelling the behavior of reactive and distributed systems [28]. Formally, an I/O automaton $A$ is a tuple $\langle states, start, sig, trans \rangle$, where:

- *states* is a finite or infinite set of *states*;
- *start* $\subseteq$ *states* is a set of *initial states*;
- *sig* is the set of *actions* of $A$ partitioned into input actions *in*, internal actions *int* and output actions *out*.
- *trans* $\subseteq$ *states* $\times$ *sig* $\times$ *states* is a set of *transitions* such that for every state $s \in states$ and every input action $\pi \in in$, there is a transition $(s, \pi, s') \in trans$.

Input and output actions enable the communication between the automaton and the environment: the environment controls input actions, while the automaton controls the output (and internal) actions. For any state $s$ and action $\pi$, if I/O automaton $A$ has some transitions of the form $(s, \pi, s')$, then $\pi$ is said to be enabled in $s$. Since an I/O automaton is unable to block any input, input actions in set *in* should be enabled in every state.

---

[3] Android apps are composed of multiple components called activities

## 3  Motivating Example

This section presents a motivating example that is also a case of interference that we discovered in our evaluation. It consists of two enforcers that work correctly when used individually, but that interfere when activated simultaneously. The two enforcers implement different usage policies for the Android `MediaPlayer` API.

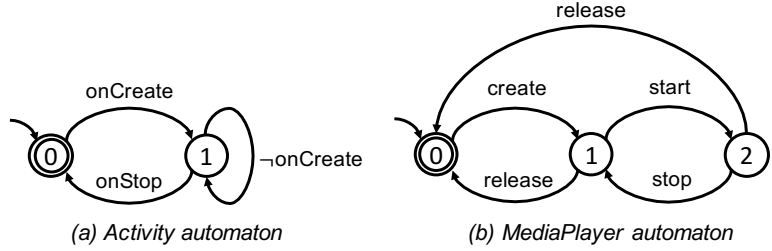

*(a) Activity automaton*  *(b) MediaPlayer automaton*

**Fig. 2.** System Automata for the `MediaPlayer` Example.

In order to describe the two enforcers, we also need to specify the behavior of a generic Android application, in terms of the lifecycle events, and the usage protocol of the `MediaPlayer` API. Note that these two elements are invariant for every application, that is, regardless of the application that is executed at runtime, the lifecycle events and the usage protocol of a `MediaPlayer` are always the same.

To keep the example small and simple, we only represent the actions that are relevant to the policies that we want to enforce. Figure 2 (a) shows the model of the Android activity lifecycle [4] limited to the creation and the stopping of an activity. Figure 2 (b) shows the usage protocol for the `MediaPlayer` API derived from the Android specifications [2].

In this example, we consider the enforcement of two policies extracted from technical and scientific documentation about the `MediaPlayer` API. The first policy is *Policy 1* introduced in Section 2, while the second policy about stopping the execution of the player is the following one:

> Policy 2: "*if you started a `MediaPlayer` and your activity receives a call to `onStop`, you must stop the `MediaPlayer`.*" [37]

The edit automaton $EA_{p1}$ that can enforce *Policy 1* is shown in Figure 1 and has been discussed in Section 2.

Figure 3 shows $EA_{p2}$, the edit automaton that can enforce *Policy 2*. As long as the `MediaPlayer` is started after the activity has been created and is stopped before the activity is stopped, the enforcer does not change the execution. However, if the activity is stopped without first stopping the `MediaPlayer` (transition from state 2 to state 0), the enforcer changes the execution inserting the `stop` action, before the execution of `onStop`.

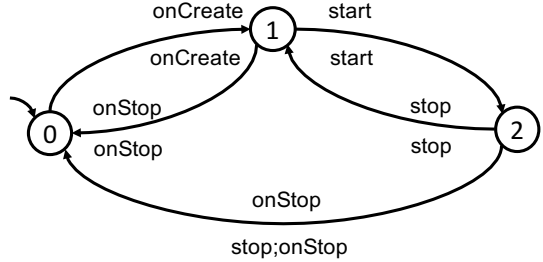

**Fig. 3.** Edit Automaton $EA_{p2}$ enforcing `Policy 2`

The enforcers in Figures 1 and 3 can interfere if they are both in their respective states 2 and the Android framework produces the `onStop` callback. In this case, both enforcers capture the `onStop` callback and attempt to change the execution. The interference occurs when the enforcer for *Policy 1* changes the execution before the enforcer for *Policy 2*.

In particular, if the automaton enforcing *Policy 1* outputs the `release` of the `MediaPlayer` instance before the other enforcer outputs the `stop` action, the `MediaPlayer` instance will be released, and the system will reach a deadlock state. In fact, the enforcer for *Policy 2* is no longer able to invoke the `stop` operation of the `MediaPlayer` instance because this call is not accepted by the `MediaPlayer` API protocol as shown in Figure 2 (b). Since the model of the resource forbids to call method *stop* from state 0, the interference results in a deadlock at the level of the models. In practice, the call to *stop* is issued by the enforcer and the execution fails due to an exception produced by the resource.

In the next section, we show how this conflict can be detected in advance and then eliminated by the developers.

## 4   Interference Analysis

The goal of the interference analysis is to check whether a set of policy enforcers can jointly operate without causing any interference. An *interference* occurs when two or more enforcers are no longer able to enforce the policies that they can enforce individually. More formally, let us assume that $Enf_1$ and $Enf_2$ are two enforcers that can operate in environment $Env$ to enforce policies $Policy_1$ and $Policy_2$, respectively. We write $Env||Enf_1 \models Policy_1$ and $Env||Enf_2 \models Policy_2$. The two enforcers $Enf_1$ and $Enf_2$ *interfere* if $Env||(Enf_1||Enf_2) \not\models Policy_1$ or $Env||(Enf_1||Enf_2) \not\models Policy_2$ or $Env||(Enf_1||Enf_2)$ includes deadlocks. This is exactly the case of the motivating example where the enforcer for

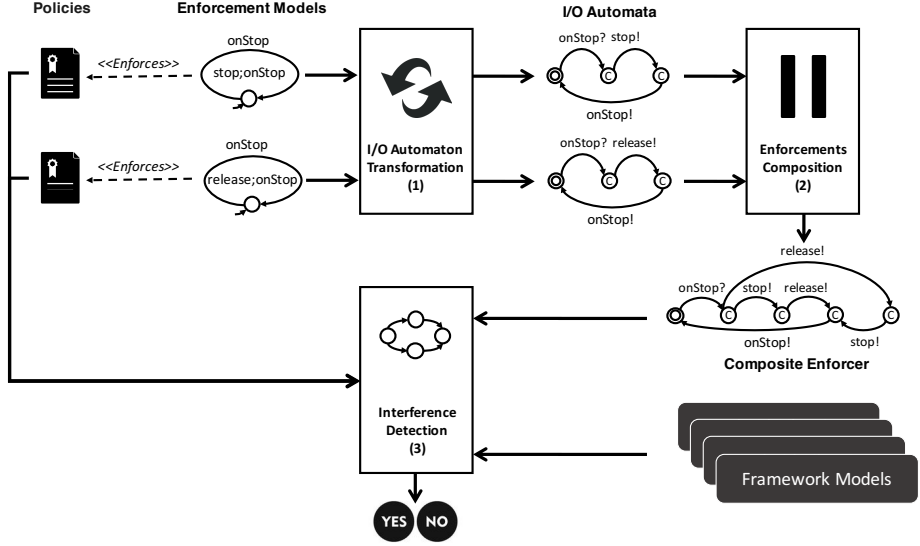

**Fig. 4.** Interference Analysis.

*Policy 1* can release the `MediaPlayer` instance before the enforcer for *Policy 2* stops the player causing a deadlock.

In our setting, the environment consists of an Android app that uses multiple resources. We represent the generic behavior of an app and the resources using one model for the app and one model for each resource, as done for the example in Figure 2. We call these models the *framework models*. Note that although we first experienced this solution in the Android environment, it is indeed valid in any environment where applications must obey to a pre-defined lifecycle and resources must be used according to a protocol, as it happens in many frameworks for the development of Web and server-side applications.

Figure 4 shows the overall structure of the interference analysis that starting from a set of *enforcement models*, the corresponding *policies*, and a set of *framework models* verifies whether the enforcers can coexist to enforce the policies without causing interferences. Since the enforcers, the apps, and the resources are communicating components, we run our analysis representing the behavior of each component as an I/O automaton. Since I/O automata provide good expressive power and a flexible framework for modeling the behavior of communicating components, they are also able to precisely capture the behavior of the components involved in real-world enforcement tasks. We thus first map the enforcers, specified as edit automata, into their corresponding I/O automata (see the *I/O automaton transformation* step in Figure 4). We then compose the enforcers to derive the *composite enforcer*, which is a single model that encapsulates the collective behavior of all the enforcers considered in the analysis (see the *enforcers composition* step in Figure 4). To check for interferences, the analysis composes the composite enforcer with the framework models and checks for the satisfaction of the policies and for the absence of deadlocks on the resulting model (see the *interference detection* step in Figure 4).

In the rest of this section, we describe these three steps in details.

### 4.1   I/O Automaton Transformation

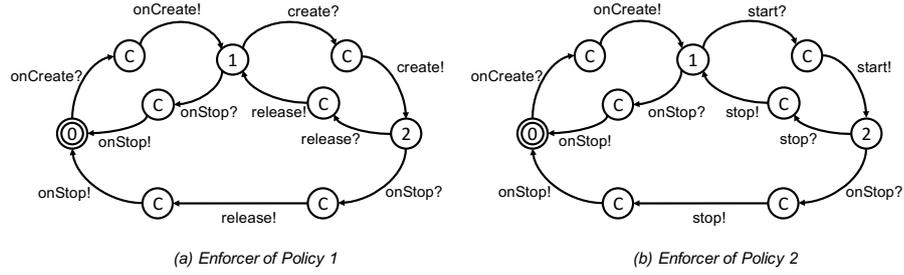

**Fig. 5.** I/O Automaton Transformation of the Running Example: a) I/O automaton $IOA_{p1}$ for *Policy 1*, b) I/O automaton $IOA_{p2}$ for *Policy 2*.

In this step, each model of enforcer encoded as an edit automaton $A_E = \langle \Sigma, Q, q_0, \delta \rangle$ is transformed into the corresponding I/O automaton $A = \langle states,$ $start, sig, trans \rangle$ according to the strategy defined below.

Since each transition in an edit automata can accept an action and produce multiple actions in response, this same behavior requires multiple transitions, and thus multiple states, to be represented in an I/O automaton. To this end, we define *states* as the union of the *origStates*, which are the same states than the states in the edit automaton, and the *newStates*, which are the additional states introduced in the I/O automaton to produce sequences of actions consistently with the transitions in the edit automaton. Since the need of these intermediate states depends on the length of the sequences that are emitted by each transition of the edit automaton, we directly exploit these sequences in the representation of the states. More formally, $states = origStates \cup newStates$, where:

- $origStates = \{\langle q, \epsilon \rangle \mid q \in Q\}$,
- $newStates = \{\langle q, s \rangle \mid q \in Q, s = a; \sigma[...i], a \in in, \delta(q, a) = \langle q', \sigma \rangle, i < |\sigma|\}$,

To preserve the intuition that these sequences of actions should be emitted quickly in response to an input, we define all the states in *newStates* as committed states. The initial state is the same as the one of the edit automaton, thus $start = \{\langle q_0, \epsilon \rangle\}$.

The operations that can be performed by the edit automaton are duplicated into input and output operations. Actually, whether an operation is an input or an output depends on whether it is accepted or emitted by a transition in the edit automaton. More formally, $sig = in \cup int \cup out$, where: $in = \{a? \mid a \in \Sigma\}$, $int = \{\}$, $out = \{a! \mid a \in \Sigma\}$.

We distinguish two main cases for the transitions. When the transition in the edit automaton suppresses the action, that is, no action is emitted, there is no need of introducing additional states in the I/O automaton to map the transition. Otherwise, extra states and transitions are needed. Formally, $trans = suppression \cup insertion$, with $suppression = \{\langle\langle q, \epsilon\rangle, a, \langle q', \epsilon\rangle\rangle \mid q, q' \in Q, a \in in, \delta(q, a) = \langle q', \epsilon\rangle\}$. In the case of $insertion$, the transitions in the edit automaton requires multiple transitions in the I/O automaton to be represented correctly. We thus distinguish three kinds of transitions that may occur in $insertion$: the first transition of a sequence, that is a transition that starts from a state in $origStates$ and reaches a state in $newStates$, an intermediate transition of a sequence, that is a transition that starts from a state in $newStates$ and reaches a state in $newStates$, and finally the last transition of a sequence, that is a transition that starts from $newStates$ and reaches a state in $origStates$. More formally, $insertion = startInsertion \cup ongoingInsertion \cup endInsertion$, where:

- $startInsertion = \{\langle\langle q, \epsilon\rangle, a, \langle q, a; \sigma[i]\rangle\rangle \mid q \in Q, a \in in, \delta(q, a) = \langle q', \sigma\rangle\}$
- $ongoingInsertion = \{\langle\langle q, a; \sigma[...i]\rangle, \sigma[i], \langle q, a; \sigma[...i + 1]\rangle\rangle \mid q \in Q, a \in in, \delta(q, a) = \langle q', \sigma\rangle, \sigma[i] \in out, 0 < i < |\sigma|\}$
- $endInsertion(A) = \{\langle\langle q, a; \sigma[...|\sigma|]\rangle, \sigma[|\sigma|], \langle q', \epsilon\rangle\rangle \mid q, q' \in Q, a \in in, \delta(q, a) = \langle q', \sigma\rangle, \sigma[|\sigma|] \in out\}$.

Figure 5 shows the output of the I/O automaton transformation step applied to the running example. Figures 5 (a) and 5 (b) show the I/O automata derived from the enforcement models for *Policy 1* and *Policy 2*, respectively. The numbered states are in $origStates$ and the numbering is consistent with the states in the original edit automaton. States marked with $c$ are the committed states in $newStates$.

The equivalence between the languages accepted by original edit automaton and the corresponding I/O automaton is pretty straightforward. By construction, every transition $t$ in the edit automaton has a corresponding linear sequence of transitions that starts by accepting an input action and continue producing the output actions consistently with $t$, and viceversa. These sequences are also linked to $origStates$ consistently with the edit automaton and the initial state is also preserved.

The only difference that the I/O automaton introduces with respect to the corresponding edit automaton is in the composition of multiple models. In fact, an output sequence emitted by the edit automaton in response to an event is atomic, while the corresponding sequence emitted through multiple states and transitions in the I/O automaton could be interrupted, although the presence of the committed states guarantee that this may happen only from another committed state. This difference is desirable in our context since the atomicity of the sequence could not be guaranteed in practice and the behavior of the enforcers should be verified without considering this property, as we do by running our analysis on the I/O automaton derived from the edit automaton.

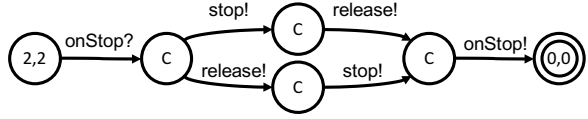

**Fig. 6.** Excerpt of Composite Automaton

### 4.2 Enforcers Composition

This step derives a composite enforcer which represents the collective behavior of all the enforcers. Since the behavior of the enforcers must be synchronized, the interference analysis derives the composite automaton using CSP (Communicating Sequential Processes)-like synchronization [21]. Thus the states of the composite automaton are the cartesian product of the states of the composed automata and its behavior is the interleaving of the behaviors of the composed I/O automata.

Considering the I/O automata derived in the motivating example (Figure 5), the state space of the resulting composite I/O automaton $CA_{p1,p2}$ is represented by pairs $\langle s_1, s_2 \rangle$, where $s_1$ is a state of I/O automaton $IOA_{p1}$, and $s_2$ is a state of I/O automaton $IOA_{p2}$. Figure 6 shows the portion of the $CA_{p1,p2}$ responsible for the interference. When the `onStop?` action is executed and $CA_{p1,p2}$ is in state $\langle 2,2 \rangle$ the model can produce both the sequences `stop!;release!` and `release!stop!`. If `release!stop!` is produced, the policies are not enforced correctly.

### 4.3 Interference Detection

This step verifies that all the policies are correctly enforced by the composite enforcer without introducing any deadlock in the system. To this end, the analysis reconstructs the global behavior of the system by composing the composite enforcer with both the framework models (i.e., the generic model of an app lifecycle and the models of the used resources) and an environment model which is simply used to generate every possible combination of actions that the app and the resources can produce (i.e., this model is used to consider every possible execution scenario in the analysis).

In this case, the composed models are I/O automata that communicate using binary synchronization channels which let pairs of automata synchronize on shared input-output actions (e.g., the output action $a!$ with the input action $a?$). Since every action emitted by the environment must be first intercepted by the composite enforcer, which reacts by generating the actions for the app and the resources, the analysis automatically renames actions to reflect the way components communicate in practice. In particular, the action produced by the environment and the corresponding actions in the framework models are renamed adding different suffixes (e.g., the `onCreate()` method is changed into `onCreate-env()!` when emitted from the environment, and into

`onCreate-app()?` when received by the app). This simple strategy prevents direct communication between the environment and the framework models. The actions in the composite enforcer are renamed to receive actions from the environment and emit actions for the app and the resources. For instance, if a transition of the enforcer receives `onCreate()?` and the following transition emits `onCreate()!`, the actions are renamed into `onCreate-env()?`, to receive the action from the environment, and `onCreate-app()!` to propagate the action to the app. This simple renaming strategy is sufficient to fully model a communication mediated by the enforcers.

To check if the enforcers can interfere, the interference analysis checks all the enforced policies specified as Computation Tree Logic (CTL)-like formulas [13] on the model resulting from the composition of the framework models, the composite enforcer, and the environment model after renaming (of course the name of the actions in the CTL formulas are renamed consistently with the model). An interference is detected if the model checker, in our case UPAAL [10], reports a counterexample that violates any policy, or the system may reach a deadlock state.

In the running example, an interference is detected because the system may reach a deadlock state. Indeed, the resulting automaton cannot proceed with the execution because the `stop` action cannot be executed on the `MediaPlayer` once it has been released. The analysis identifies the problem as a deadlock because the model of the resource does not allow the execution of `stop` after `release`. In practice, the enforcer anyway tries to invoke the `stop` method on the resource and the execution fails due to an exception returned by the resource.

## 5   Analysis of Resource Usage Policies in Android

To evaluate the effectiveness of the interference analysis, we focus on misuses of the APIs that provide access to critical system resources, such as camera and the media player. Misuses of these APIs are frequent in Android [20, 37] and often cause resource leaks which lead to performance degradation and crashes.

To identify the correctness policies that can be enforced on Android apps that interact with system resources, we exploited the recommendations about API usage derived from the Android documentation [1, 3] by Wu et al. [37]. We identified ten different resources that must satisfy multiple policies, for a total of twenty-five policies. We encoded each policy as a CTL formula and defined the corresponding enforcer. We finally used the interference analysis presented in this paper to detect interferences between enforcers, that is, enforcers that work well in isolation but fail to enforce the policies when used jointly with other enforcers.

Table 1 shows the obtained results. Column **API** indicates the API that provides access to a specific resource. Below the name of the API we report the name of its package. Column *Resource Usage Policy* lists the set of policies that each API must satisfy. We have written the policies in the form "`<acquire method>`/`<release method>`: `<callback>`" which should be interpreted as:

**Table 1.** Interference Analysis of Resource Usage Policies

| API | Resource Usage Policy | Interference |
|---|---|---|
| BluetoothAdapter (android.bluetooth) | enable/disable: onDestroy<br>startDiscovery/cancelDiscovery: onDestroy<br>getProfielProxy/closeProfileProxy: onDestroy | No |
| Camera (android.hardware) | lock/unlock: onPause<br>open/release: onPause<br>startFaceDetection/stopFaceDetection: onPause<br>startPreview/stopPreview:onPause | **Yes** |
| AudioManager (android.media) | requestAudioFocus/abandonAudioFocus: onPause<br>startBluetoothSco/stopBluetoothSco: onPause<br>loadSoundEffects/unloadSoundEffects: onPause | No |
| MediaCodec (android.media) | createDecoderByType/release: onPause<br>start/stop: onPause | **Yes** |
| MediaPlayer (android.media) | <init>/release: onStop<br>create/release: onStop<br>start/stop: onStop | **Yes** |
| MediaRecorder (android.media) | <init>/release: onStop<br>start/stop: onStop | **Yes** |
| NfcAdapter (android.nfc) | enableForegroundDispatch/<br>disableForegroundDispatch: onPause<br>enableForegroundNdefPush/<br>disableForegroundNdefPush: onPause | No |
| RemoteCallbackList (android.os) | beginBroadcast/finishBroadcast: onDestroy<br>register/unregister: onDestroy | No |
| Surface (android.view) | <init>/release: onDestroy<br>lockCanvas/unlockCanvasAndPost: onDestroy | **Yes** |
| SurfaceHolder (android.view) | addCallback/removeCallback: onDestroy<br>LockCanvas/unlockCanvasAndPost: onDestroy | No |

if the app invokes `<acquire method>`, it should also invoke `<release method>` when a call to `<callback>` is received, unless `<release method>` has been already invoked before. Column *Interference* indicates the result of the interference analysis of the enforcers that enforce the specified policies: *No* indicates that the enforcers combined together are still able to successfully enforce all the policies, while *Yes* indicates that an interference among the enforcers has been detected.

In order to observe the impact that interferences have on the actual execution of an app, we have implemented and deployed the analyzed enforcers on a real device as described in [32]. After activating the interfering enforcers and opening an app that violates the policy, we execute a test case that reproduces the scenario with the misuse and we observed that in all the cases interference caused the crash of the app.

Interestingly, we reported an interference for 5 out of 10 analyzed APIs. This result shows that interference among policy enforcers can be a major obstacle to the successful deployment of the policy enforcers technology. The mechanism presented in this paper can be a useful tool to avoid these situations. It can be used to decide which enforcers to activate and which enforcers to not activate. For instance, it is not possible to activate the four enforcers specified for the `Camera` API, but our analysis reveals that the first and third enforcers of the camera are compatible and thus can be activated together. Moreover, the developers can exploit this result to redesign some of the enforcers.

This result also suggests that sets of enforcers cannot be designed in a completely independent way, but their co-existence must be planned in advance and reflected in their definition. In this paper we do not discuss how to evolve enforces in this direction, we left this research direction for future work.

## 6   Related Work

Runtime solutions for avoiding and mitigating the impact of failures have been studied in many different contexts, including Web applications [29], mobile applications [31, 32, 16, 15], operative systems [35], and Cloud environments [14].

In the context of the Android environment, runtime enforcement mechanisms have been focused on the enforcement of privacy [16] and resource usage policies [32], obtained respectively by applying mechanisms for detecting and disabling suspicious method calls, and by augmenting classic Android libraries with proactive mechanisms able to automatically suppress and insert API calls. Both approaches are not intrinsically limited to security and resource usage policies, but could be potentially exploited to generally enforce correctness policies.

So far, these approaches focused on the definition of the enforcement mechanisms and paid little attention to the interference between mechanisms, which might be an issue when multiple policies must be enforced. The work presented in this paper is complementary to these approaches because it provides an analysis framework for checking the compatibility between enforcers.

The problem of handling interferences has been considered in the work by Bauer et al. [9]. In their work, Bauer et al. present a framework that can be used by the developers to specify how the enforcement mechanism should behave when multiple enforcers directly interfere, that is, multiple enforcers try to alter an execution as a reaction to a same action. To address these situations, Bauer et al. [9] define several composition operators that can be used to obtain a strategy to solve these situations. General composition operators for enforcers, which might be potentially used to reason on interferences, have been also defined by Falcone et al. [19]. Compared to these strategies, the analysis presented in this paper can address a broader set of situations, not only the direct inference. Moreover, it can also be exploited to know a-priori if a set of enforcers are compatible, instead of lately discovering it at run-time, once their interference or the lack of application of some enforcers may have serious consequences for the health of the system.

A body of work formally studied the classes of properties that can be enforced using different models and languages, with an emphasis on security policies [33, 26, 27, 17, 23]. Interestingly these approaches should be complemented with appropriate analysis routines to check that the result of the enforcement is in line with what the enforcers are expected to achieve. The gap between the policies to be enforced and the enforced behaviors has been highlighted by Bielova et al [11] who show that often there is little guarantee that enforcers fix the bad sequences in the desired way. This result further stresses the need of analysis strategies similar to the one presented in this paper.

## 7   Conclusions and Future Work

*Conclusions.* The reliability of software applications can be improved by exploiting advanced execution environments equipped with mechanisms to enforce correctness policies, such as security [16, 27] and resource usage policies [32, 16]. Although enforcers can be effective when used in isolation, their effect on the application and the execution environment when executed jointly might be hard to predict and potentially harmful. In particular, a set of enforcers may fail to enforce the policies that they are designed to enforce individually.

To address this problem, we presented an analysis framework that can be used to detect interferences among enforcers. The analysis can be exploited by both the developers, to improve the enforcement strategies and implement enforcers that can safely co-exist, and the users, to identify sets of policies that can be enforced without introducing side-effects.

Our initial evaluation with several enforcers designed to guarantee the correct usage of multiple Android resources revealed that enforcers may easily interfere. This result suggests that defining techniques to design interference-free enforcers, as well as defining efficient and effective verification mechanisms, are important challenges for the future.

*Future work.* In this work we present a possible analysis, but there are several complementary aspects worth to be analyzed. For instance, timing aspects in runtime enforcement [30, 18] have not been considered, but timing could be another source of interferences. For instance, the joint activation of two enforcers may successfully cause the enforcement of some security policies, but may cause serious slow downs that dramatically annoy users.

Finally, while in this work we focused on revealing interferences, we plan to investigate mechanisms to semi-automatically or automatically fix interferences.

**Acknowledgment** This work has been partially supported by the H2020 Learn project, which has been funded under the ERC Consolidator Grant 2014 program (ERC Grant Agreement n. 646867), the GAUSS national research project, which has been funded by the MIUR under the PRIN 2015 program (Contract 2015KWREMX), and the COST Action ARVI IC1402, supported by COST (European Cooperation in Science and Technology).